\documentclass[showpacs, twocolumn, amsmath, amsfonts, prb]{revtex4}
\usepackage{graphicx}
\usepackage{dcolumn}
\usepackage{amsmath}
\usepackage{amssymb}

\begin{document}
\title{Magnetization suppression of Type-II Superconductors by external alternating
magnetic field}
\author{Alex Levchenko} \affiliation{Department of
Physics, University of Minnesota, Minneapolis, MN 55455, USA}
\begin{abstract}
The effect of suppression of static magnetization of an
anisotropic hard superconductor by alternating magnetic field is
analyzed theoretically. The magnetic moment suppression dynamics
is described with respect to the magnetization loop of the
superconductor. It is found that in some cases the magnetic moment
varies nonmonotonically with the growth in amplitude $\mathrm{h}$
of the alternating field. Effect of transition, induced by
$\mathrm{\mathbf{h}}(t)$, of superconductor form paramagnetic into
the diamagnetic state is considered. The amplitude of alternating
magnetic field $\mathrm{h}_{c}(\delta,\vartheta)$ for which the
complete suppression of the magnetization occurs is calculated as
the function of anisotropy parameter $\delta$ and it orientation
angle $\vartheta$ with respect to the crystallographic axes of the
sample.
\end{abstract}
 \maketitle
\section{Introduction}
For many years the physics of the vortex matter in the
superconductors attracts the attention of researchers. This is
because collective properties of the vortices reveal variety of
reach and very interesting phenomena which involve different
phases and phase transitions
Ref.\cite{Larkin,Phase1,Phase2,Phase3,Phase4}, magnetic
instabilities Ref.\cite{Instabil1,Instabil2,Instabil3,Instabil4}
mesoscopic and fluctuation effects and others. In this paper one
example of the instability effects, namely the problem of magnetic
moment suppression by external alternating magnetic filed, is
considered in some details. Usually the static and quasi-static
electromagnetic properties of hard superconductors is treated in
terms of critical state model at first suggested by
Bean~\cite{Bean}. This model describes of magnetic induction $B$
distribution inside the superconductors. In the simplest possible
case, when superconductor represents rectangular slab with x-axis
directed perpendicular to it planes and z-axes along the magnetic
field direction, the Bean equation has following form $\partial
B/\partial x=\pm 4\pi J_{c}/c$, here $J_{c}$ is critical current
density. Critical state equation can be easily understood by
analyzing the forces balance which acts on a single vortex. In
accordance to the model, the Lorentz force, which acts on the
vortex, is compensated by the pinning forces between the vortex
and various defects in the crystalline lattice. Later critical
state model was generalized for more complicated situations in
several directions.

First generalization was performed by the authors of
Ref.~\cite{Fisher} in order to include quasi-stationary processes.
In the most of the cases, for simple geometries, direction of the
critical current vector $\mathbf{J}_{c}$ is uniquely determined.
But situation is some what more complicated when external magnetic
field has several spatial components or when its varies in time.
For these time dependent cases usually critical state model is
written as $\mathrm{curl}\mathbf{B}=(4\pi J_{c}/c)\mathbf{E}/E$,
where $\mathbf{E}$ is electric field. This model assumes that
varying in time magnetic flux produces electric field,
superconductor becomes in the resistive state and the direction of
electric current coincide with the direction of $\mathbf{E}$
similar to normal metal. It is also important to answer the
question about the direction of critical current density in that
regions of superconductor where there is no electric field. At
this point rises one of the most important electrodynamic
properties of hard superconductors. The magnetic state of the
superconducting sample, at some specific moment of the time, is
described not only by critical model equations with some boundary
conditions but also depends on prehistory - how this state was
prepared. In order to understand this statement lets consider
example of static magnetization. Assume that we have
superconducting slab placed in the external magnetic field with
the quasi-stationary varied amplitude. Because magnetic field
changes in time very slowly we can assume that in each instance
superconducting sample is in the critical state with the critical
current density $\mathbf{J}_{c}(B)$ which is determined by the
value of magnetic induction at this moment of the time. Direction
of this current coincides with the direction of electric field
$\mathbf{E}$ which emerges from slowly varying external magnetic
field $\mathbf{H}$. This picture is valid while amplitude of
$\mathbf{H}$ changes. As soon as magnetic field stop changes the
electric field disappears but critical current density persists
its direction which was set by the electric field at last instance
of its existence. Exactly this dissipationless current determines
final magnetization of the sample which is usually called static.

It is also important to point out that critical state model
equations are significantly nonlinear and this nonlinearity is
peculiar only for superconductors and has no equivalents in other
nonlinear medias. This specific nonlinearity leads to several very
interesting effects\cite{MagnLoop} one of which is static
magnetization suppression. The essence of this effect is the
following. Let a plane superconducting sample cooled in zero
magnetic field be placed into an external magnetic field
$H>H_{c1}$ (here $H_{c1}$ is the lower critical magnetic field)
which is parallel to the superconductor surface. Pinning leads to
the emergence of a nonuniform distribution of the magnetic
induction and, accordingly, a static magnetization, in the sample.
If an alternating magnetic field
$\mathrm{\mathbf{h}}(t)=\mathrm{\mathbf{h}}\cos(\omega t)$ is
applied to the magnetized sample in a direction parallel to the
sample surface and perpendicular to a constant field $\mathbf{H}$,
then magnetization M of the sample decreases. Direct
measurements~\cite{Voloshin} show that everywhere in the sample
where the alternating magnetic field penetrates, the flow of
nondissipative currents becomes impossible. The current that
previously screened field $\mathbf{H}$ and contributed to the
magnetization of superconductor disappear. If the amplitude is
sufficiently large $h\sim H_{p}=2\pi J_{c}(H)/cd$ (the alternating
field is practically penetrate through the entire sample)
magnetization is suppressed.

Second generalization of critical state model is related to the
fact of high anisotropy of the high-$T_{c}$ superconductors which
will sufficiently important in this work. In the anisotropic model
critical current density becomes a tensor $J_{cij}$ and
straightforward generalization of model equations gives
$(\mathrm{curl}B)_{i}=(4\pi/c)J_{cij}E_{j}/E$. Diagonal components
of the $J_{cij}$ tensor are significantly different, for example
for $\mathrm{YBaCuO}$ the critical current density in the
$\mathbf{ab}$ plane is much grater the that in $\mathbf{c}$
direction.

In this work the theoretical analysis of static magnetization
suppression in the anisotropic case under influence of alternating
magnetic field $\mathbf{h}(t)=\mathbf{h}\cos(\omega t)$ is
performed. It is known that magnetization curve $M(H)$ of HTSC
reveals hysteresis and it turns out that character of
magnetization suppression is very sensitive to the position of
current magnetization in the magnetization loop (this reflection
of the sensitivity to the magnetic prehistory described above). It
is found that in total there are nine regions in the magnetization
loop where suppression scenario is qualitatively different and
dynamics of magnetic moment suppression with growth of the
amplitude $\mathrm{h}$ is described in each of these regions.

\section{Main equations and geometry of the problem}

Consider an infinite plane-parallel superconducting plate of
thickness $d$ placed in external constant magnetic field
$\mathbf{H}$ and alternating magnetic field
$\mathbf{h}(t)=\mathbf{h}\cos(\omega t)$ which are mutually
perpendicular and parallel to the plate surface. It  is assumed
that all fields and currents depend on only one spatial coordinate
$x$ directed along the normal to the plate. The origin $x=0$ is
located in the origin of the sample. In this geometry the
equations of generalized critical state model for magnetic
induction $\mathbf{B}$, written in the components, take form
\begin{subequations}\label{dcMagneticInduction}
\begin{equation}
\frac{\partial B_{z}}{\partial
x}=-\frac{4\pi}{c}J_{cy}(B_{y},B_{z})\cos(\varphi(x)),
\end{equation}
\begin{equation}
\frac{\partial B_{y}}{\partial
x}=\frac{4\pi}{c}J_{cz}(B_{y},B_{z})\sin(\varphi(x)),
\end{equation}
\end{subequations}
here $\varphi(x)$ - angle between the electric field vector
$\mathbf{E}$ and $y$ axis. The spatially averaged component of the
magnetization $\mathbf{M}$ along the direction of the external
magnetic field $\mathbf{H}$, which we will denote as $M_{H}$, is
given by following formula
\begin{equation}\label{Magnetization}
M_{H}=\frac{1}{4\pi}\left[\frac{1}{d}\int_{-d/2}^{d/2}(B_{z}(x)\cos\vartheta+
B_{y}(x)\sin\vartheta)\mathrm{d}x-H\right]
\end{equation}
Eqs.~(\ref{dcMagneticInduction}) and (\ref{Magnetization}) should
be accompanied by the Maxwell equations for electric field
\begin{equation}
\frac{\partial E_{z}}{\partial x}=\frac{1}{c}\frac{\partial
\mathrm{h}_{y}}{\partial t},\quad\frac{\partial E_{y}}{\partial
x}=-\frac{1}{c}\frac{\partial \mathrm{h}_{z}}{\partial t}
\end{equation}
and boundary conditions
\begin{equation}
B_{z}\left(\pm\frac{d}{2}\right)=H\cos\vartheta,\quad
B_{y}\left(\pm\frac{d}{2}\right)=H\sin\vartheta.
\end{equation}
It is clear that exact integration of
Eqs.(\ref{dcMagneticInduction}) is very complicated problem due to
specific nonlinearity of equations. For the sake of simplifying of
the calculations, the dependence of the components
$J_{c}(\mathbf{B}(x))$ of the critical current density on the $x$
coordinate, caused by the nonuniformity of the magnetic induction
distribution, will be neglected and assumed that
$J_{c}(\mathbf{B}(x))=J_{c}(\mathbf{H})$. In addition, the $x$ -
coordinate dependent angle $\varphi(x)$ between the electric field
and the $y$ axis will be replaced by the angle $\pi/2+\vartheta$
between the external alternating magnetic field and the $z$ axis
(as in the anisotropic situation). It turns out that these
simplifications have no qualitative influence on the results but
make it possible to perform analytical calculations completely.

For further convenience lets introduce following dimensionless
variables
\begin{equation}
\xi=\frac{2x}{d},\quad\mathcal{H}=\frac{H}{H_{p}},\quad\mathcal{B}=\frac{B}{H_{p}},\quad
H_{p}=\frac{2\pi dJ_{cy}}{c},
\end{equation}
\[
h=\frac{\mathrm{h}}{H_{p}},\quad b=\frac{\mathrm{b}}{H_{p}}
,\quad\mathcal{M}=\frac{M_{H}}{H_{p}},\quad\delta=\frac{J_{cz}}{J_{cy}}
\]
here capital letters refer to the dc-magnetic field $\mathbf{H}$
and small letters to the ac-magnetic field $\mathbf{h}(t)$.
\section{Calculation of Magnetic moment $\mathcal{M}$ suppression dynamics}

The main idea of calculations is as follows, first of all we have
to find static magnetization when the alternating magnetic field
is absent. This part of the problem is well developed both
theoretically and experimentally and we can refer to
Ref.~\cite{MagnLoop} where magnetization loop was studied in
details. For each point of the magnetization loop there is some
specific distribution of magnetic induction which determines
magnetization. As the second step we have to solve dynamical
problem and describe the penetration of ac-magnetic filed. In this
way it will be possible to find the penetration depth $\xi_{h}$
and critical amplitude of ac-field $h_{c}(\vartheta)$ at which
total suppression of magnetization occurs. As soon as problem is
two-dimensional it means that from two penetration-depths of each
spatial component of alternating magnetic field we have to choose
the highest.

One can easy shows that at all places where the alternating field
$h(t)$ penetrates, the magnetic induction can be presented as a
sum of two terms. One of these terms is a constant homogeneous
quantity coinciding with vector $\mathbf{\mathcal{H}}$. The second
term, which corresponds to the nonhomogeneous magnetic induction
distribution of ac-magnetic field $b_{y,z}$, may be described by
the following equations (in analogy with
Eqs.(\ref{dcMagneticInduction}))
\begin{equation}\label{acMagnDistrZY}
\frac{\partial b_{z}}{\partial
\xi}=\sin\vartheta,\quad\frac{\partial b_{y}}{\partial
\xi}=\delta\cos\vartheta.
\end{equation}
These equations hold in the superconductor region where both
induction components $b_{z}$ and $b_{y}$ are present. In the
region there component $b_{z}$ vanishes, which correspond to
$\varphi(x)=\pi/2$ or equivalently $\vartheta=0$, and there is
only induction component $b_{y}$, the distribution of this
component is described by the equation
\begin{equation}\label{acMagnDistrY}
\frac{\partial b_{y}}{\partial \xi}=\delta.
\end{equation}
Solution of Eqs.(\ref{acMagnDistrZY}) and Eg.(\ref{acMagnDistrY})
is easy to find
\begin{subequations}
\begin{equation}
b_{z}(\xi)=h\sin\vartheta+\sin\vartheta(\xi-1),\quad
\xi_{z}\leq\xi\leq 1,
\end{equation}
\begin{equation}
b_{y}(\xi)=\left\{ \begin{array}{cc}
h\cos\vartheta+\delta\cos\vartheta(\xi-1) & \xi_{z}\leq\xi\leq 1, \\
b_{y}(\xi_{z})+\delta(\xi-\xi_{z}) & \xi_{y}\leq\xi\leq \xi_{z},
\end{array}\right.
\end{equation}
\end{subequations}
here $\xi_{z}$ and $\xi_{y}$ are penetration depths of each
component of ac-magnetic field. Cusp in the distribution of
$b_{y}$ component is the result of the fact that coordinate
dependence of angle $\varphi(x)$ was neglected. From the condition
$b_{z}(\xi_{z})=0$ one finds the penetration depth for $z$
component of ac-magnetic field $\xi_{z}=1-h$ and value of $b_{y}$
at that point $b_{y}(\xi_{z})=(1-\delta) h\cos\vartheta$.
Similarly one can find penetration depth for y-component
$\xi_{y}=1-h-\left(\frac{1-\delta}{\delta}\right)h\cos\vartheta$
and finally from the condition $b_{y}(\xi=0)=0$ critical amplitude
$h_{c}(\vartheta)$ of the alternating magnetic field at which
total suppression of magnetization occurs
\begin{equation}\label{CriticalField}
h_{c}(\vartheta)=\frac{\delta}{\delta+(1-\delta)\cos\vartheta}.
\end{equation}
Function $h_{c}(\vartheta)$ has universal character; it is defined
via the anisotropy parameter $\delta$ of the theory which makes
this formula to be interesting from the point of view of
experiment. Say by measuring the $h_{c}(\vartheta)$ at
$\vartheta=0$ one can directly find anisotropy parameter because
$\delta=h_{c}$ or $\delta=ch_{c}/2\pi dJ_{cy}$ in dimension
variables.

Let now consider the dynamics of suppression of the static
magnetic moment $\mathcal{M}$ of the sample by an orthogonal
alternating magnetic field. As has been noted above, the
suppression effect essentially depends on the magnetic prehistory
of the sample, i.e. on the position of the starting point on the
magnetization loop. Below will be considered the most simple but
nevertheless practically interesting case of the magnetic
prehistory, when the external magnetic field monotonically
increased up to certain maximal value $\mathcal{H}_{m}$, such that
$1\ll\mathcal{H}_{m}\ll\mathcal{H}_{c2}$, and then decreased back
to zero. We will distinguish between two cases namely
\textit{direct magnetization}, when $\mathcal{H}$ monotonically
increased $\mathcal{H}\in[0,\mathcal{H}_{m}]$ and \textit{reverse
magnetization} when magnetic field $\mathcal{H}$ gradually
decreased from its maximum value to zero
$\mathcal{H}\in[\mathcal{H}_{m},0]$. It turns out that there are
three distinct regions for the direct magnetization and six for
the reverse where suppression occurs qualitatively and
quantitatively different all these regions will be discussed. For
each of these regions the dependence of the magnetization on the
amplitude of alternating magnetic field $\mathcal{M}(h)$ will be
found.
\subsection{Direct Magnetization}
For the direct magnetization there are three distinct regions
$\mathcal{H}\leq h_{c}(\vartheta)$,
$h_{c}(\vartheta)\leq\mathcal{H}\leq 1$ and
$1\leq\mathcal{H}\leq\mathcal{H}_{m}$. The calculations of
$\mathcal{M}(h)$ are straightforward but cumbersome so that the
only final result will presented and discussed.

$\mathbf{A.1}$ Range: $0<\mathcal{H}\leq h_{c}(\vartheta)$
\begin{subequations}\label{Region-A-1}
\begin{equation}\label{A-1-1}
\left\{
\begin{array}{l} 0<h\leq h_{c}(\vartheta)-\mathcal{H},\\
\mathcal{M}(h)=-\frac{\mathcal{H}}{4\pi}\left(1-\frac{h}{h_{c}(\vartheta)}\right)
+\frac{\mathcal{H}^2}{8\pi}+\frac{\mathcal{H}^{2}(1-\delta)\sin^{2}\vartheta}{8\pi
h_{c}(\vartheta)},
\end{array}\right.
\end{equation}
\begin{equation}\label{A-1-2}
\left\{\begin{array}{l} h_{c}(\vartheta)-\mathcal{H}<h\leq h_{c}(\vartheta)(1-\mathcal{H}),\\
\mathcal{M}(h)=\left[\frac{(1-\delta)\mathcal{H}\sin^{2}(\vartheta)}{8\pi}+
\frac{\delta\mathcal{H}\sin\vartheta}{8\pi
h_{c}(\vartheta)}+\frac{\delta\mathcal{H}\sin\vartheta}{8\pi}-\frac{\mathcal{H}}{4\pi}\right]\times\\
\left(1-\frac{h}{h_{c}(\vartheta)}\right)+
\frac{\mathcal{H}^2}{8\pi}-\frac{\delta\mathcal{H}^2\sin\vartheta}{8\pi
h_{c}(\vartheta)}-\frac{\delta\sin\vartheta}{8\pi}\left(1-\frac{h}{h_{c}(\vartheta)}\right)^{2}
\end{array}\right.
\end{equation}
\begin{equation}\label{A-1-3}
\left\{
\begin{array}{l} h_{c}(\vartheta)(1-\mathcal{H})\leq h\leq h_{c}(\vartheta),\\
\mathcal{M}(h)=-\frac{1}{8\pi}(\cos^{2}\vartheta+\delta\sin^{2}\vartheta)
\left(1-\frac{h}{h_{c}(\vartheta)}\right)^{2}.
\end{array}\right.
\end{equation}
\end{subequations}

$\mathbf{A.2}$ Range: $h_{c}(\vartheta)<\mathcal{H}\leq 1$ -- if
amplitude of the ac-magnetic filed is at the range
$0<h<h_{c}(\vartheta)(1-\mathcal{H})$ then magnetization
$\mathcal{M}(h)$ is described by the formula Eq.(\ref{A-1-2}). For
a higher amplitudes $h_{c}(\vartheta)(1-\mathcal{H})\leq h\leq
h_{c}(\vartheta)$ the $\mathcal{M}(h)$ is given by
Eq.(\ref{A-1-3}).

$\mathbf{A.3}$ Range: $1<\mathcal{H}\leq\mathcal{H}_{m}$ -- for
any amplitude of the alternating magnetic field at the interval
$0<h\leq h_{c}(\vartheta)$ the suppression of $\mathcal{M}(h)$
occurs in accordance with the formula Eq.(\ref{A-1-3}).

Generally speaking the dynamical suppression of $\mathcal{M}$ for
the direct magnetization occurs by the same scenario for all three
regions in the magnetization loop
$\mathcal{H}\in[0,h_{c}(\vartheta)]$,
$\mathcal{H}\in[h_{c}(\vartheta),1]$ and
$\mathcal{H}\in[1,\mathcal{H}_{m}]$. Because of that we will
discuss only the case $\mathbf{A.1}$ and for two others situation
is essentially the same. Inside of the region
$\mathcal{H}\in[0,h_{c}(\vartheta)]$ we can distinguish three
steps in the evolution of the magnetic induction distribution, and
subsequently magnetization $\mathcal{M}$, in the sample. Each of
these steps occurs at some specific range of amplitudes of
alternating magnetic field which is described by inequalities in
the formulas Eqs.(\ref{A-1-1})-(\ref{A-1-3}). Static magnetization
is negative which means that sample is in the diamagnetic state.
With the gradual increase of $h$ the static magnetization is
smoothly vanishes. This happens because ac-field penetrates dipper
inside the sample, magnetic induction in that region becomes
homogeneous and doesn't contribute to the magnetization. Full
suppression occurs at the critical filed Eq.(\ref{CriticalField}).

\subsection{Reverse Magnetization}
Suppression of $\mathcal{M}$ for the reverse magnetization is
qualitatively different from that of direct magnetization. For all
magnetic filed ranges in the magnetization loop (only with
exception for the last one) there is characteristic cusp in the
distribution of the magnetic induction. This cusp is the result of
redistribution the magnetic induction after lowering amplitude of
$\mathcal{H}$ below the $\mathcal{H}_{m}$ and there is peculiar
feature of critical state model. Precisely this cusp is
responsible for some new effects. This happens because now static
magnetization has two terms one of which is negative but another
is positive. Depending on the value of $\mathcal{H}$ and $h$ one
of this terms wins such that magnetization may be either positive
(paramagnetic) or negative (diamagnetic) and
paramagnetic-diamagnetic transition is possible.

$\mathbf{B.1}$ Range:
$\mathcal{H}_{m}-\delta<\mathcal{H}<\mathcal{H}_{m}$
\begin{subequations}\label{Region-B-1}
\begin{equation}\label{B-1-1}
\left\{
\begin{array}{l} 0<h\leq h_{c}(\vartheta)(\mathcal{H}_{m}-\mathcal{H}),\\
\mathcal{M}(h)=-\frac{1}{8\pi}(\cos^{2}\vartheta+\delta\sin^{2}\vartheta)
\left(1-\frac{h^{2}}{2h^{2}_{c}(\vartheta)}\right)+\\
+\frac{\mathcal{H}_{m}-\mathcal{H}}{4\pi}\left(1-\frac{h}{2h_{c}}\right)-
\frac{(\mathcal{H}_{m}-\mathcal{H})^{2}}{16\pi\delta}(\delta\cos^{2}\vartheta+\sin^{2}\vartheta),
\end{array}\right.
\end{equation}
\begin{equation}\label{B-1-2}
\left\{\begin{array}{l} h_{c}(\mathcal{H}_{m}-\mathcal{H})\leq
h<\delta^{-1}h_{c}(\vartheta)(\mathcal{H}_{m}-\mathcal{H}),\\
\mathcal{M}(h)=\frac{(\mathcal{H}_{m}-\mathcal{H})\sin^{2}\vartheta}{4\pi}
\left(1-\frac{h}{2h_{c}(\vartheta)}\right)-\frac{(\mathcal{H}_{m}-
\mathcal{H})^{2}\sin^{2}\vartheta}{16\pi\delta}-\\
-\frac{\cos^{2}\vartheta}{8\pi}\left(1-\frac{h}{h_{c}(\vartheta)}\right)^{2}
-\frac{\delta\sin^{2}\vartheta}{8\pi}\left(1-\frac{h^{2}}{2h^{2}_{c}(\vartheta)}\right)
\end{array}\right.
\end{equation}
\end{subequations}
In the amplitude interval
$\delta^{-1}h_{c}(\vartheta)(\mathcal{H}_{m}-\mathcal{H})\leq
h\leq h_{c}(\vartheta)$ the magnetization is described by the
formula Eq.(\ref{A-1-3}).

In the case of $\mathbf{B.1}$ we again can distinguish three steps
in the magnetization dynamics. Despite the existence of the
positive term in the static magnetization the sample still
diamagnetic. As amplitude of alternating field increases sample
becomes even more diamagnetic because at the beginning the
suppression affects only positive component of the magnetization.
After positive component is suppressed completely, magnetization
reaches its minimum negative value. Further suppression occurs by
the same scenario as it was for $\mathbf{A.1}$ Eq.(\ref{A-1-3}).

$\mathbf{B.2}$ Range:
$\mathcal{H}_{m}-2\delta/(1+\delta)<\mathcal{H}\leq\mathcal{H}_{m}-\delta$.
In the interval of the amplitudes
$0<h<h_{c}(\vartheta)(\mathcal{H}_{m}-\mathcal{H})$ magnetization
is defined by the formula Eq.(\ref{B-1-1}) and in the interval
$h_{c}(\vartheta)(\mathcal{H}_{m}-\mathcal{H})\leq
h<h_{c}(\vartheta)(2-\delta^{-1}(\mathcal{H}_{m}-\mathcal{H}))$ by
the formula Eq.(\ref{B-1-2}). For the final interval we have
\begin{equation}\label{Region-B-2}
\left\{
\begin{array}{l} h_{c}(\vartheta)(2-\delta^{-1}(\mathcal{H}_{m}-\mathcal{H}))
\leq h\leq h_{c}(\vartheta),\\
\mathcal{M}(h)=-\frac{1}{8\pi}(\cos^{2}\vartheta-\delta\sin^{2}\vartheta)
\left(1-\frac{h}{h_{c}(\vartheta)}\right)^{2}.
\end{array}\right.
\end{equation}
Here dynamical picture of magnetic moment suppression is the same
as in $\mathbf{B.1}$.

$\mathbf{B.3}$ Range:
$\mathcal{H}_{m}-2\delta<\mathcal{H}\leq\mathcal{H}_{m}-2\delta/(1+\delta)$.
In the amplitude interval $0\leq
h<h_{c}(\vartheta)(2-\delta^{-1}(\mathcal{H}_{m}-\mathcal{H}))$
the $\mathcal{M}(h)$ is described by the formula Eq.(\ref{B-1-1})
and for the next interval we have
\begin{equation}\label{Region-B-3}
\left\{
\begin{array}{l} h_{c}(\vartheta)(2-\delta^{-1}(\mathcal{H}_{m}-\mathcal{H})
\leq h\leq h_{c}(\vartheta)(\mathcal{H}_{m}-\mathcal{H}),\\
\mathcal{M}(h)=\frac{(\mathcal{H}_{m}-\mathcal{H})\cos^{2}\vartheta}{4\pi}
\left(1-\frac{h}{2h_{c}(\vartheta)}\right)-
\frac{(\mathcal{H}_{m}-\mathcal{H})^2\cos^{2}\vartheta}{16\pi}
\\+\frac{\delta\sin^{2}\vartheta}{8\pi}
\left(1-\frac{h}{h_{c}(\vartheta)}\right)^2-\frac{\cos^{2}\vartheta}{8\pi}
\left(1-\frac{h^{2}}{2h^{2}_{c}(\vartheta)}\right).
\end{array}\right.
\end{equation}
And finally for the interval
$h_{c}(\vartheta)(\mathcal{H}_{m}-\mathcal{H})\leq h\leq
h_{c}(\vartheta)$ is described by Eq.(\ref{Region-B-2}).

$\mathbf{B.4}$ Range:
$\mathcal{H}_{m}-1<\mathcal{H}\leq\mathcal{H}_{m}-2\delta$. For
the case $0\leq h<h_{c}(\vartheta)(\mathcal{H}_{m}-\mathcal{H})$
for $\mathcal{M}(h)$ we have Eq.(\ref{Region-B-3}) and for the
interval $h_{c}(\vartheta)(\mathcal{H}_{m}-\mathcal{H})\leq h\leq
h_{c}(\vartheta)$ magnetization is described by the formula
Eq.(\ref{Region-B-2}).

As soon as physical processes are similar for the ranges
$\mathbf{B.3}$ and $\mathbf{B.4}$ we will discuss them together.
Because of relatively big contribution from the positive component
of the magnetic moment, sample at the beginning is paramagnetic.
Increase in the amplitude of the alternating magnetic field causes
decrease in the magnetic moment because at the beginning
suppression affects only positive component of the magnetization.
At some characteristic amplitude of ac-filed, which is smaller
then $h_{c}(\vartheta)$, magnetization becomes zero. This
situation correspond to the case when contributions from positive
and negative components to the total magnetization are equal. For
further increase of the amplitude $h$ sample becomes diamagnetic,
magnetic moment reaches its minimal value then growth back and
suppress completely at the field $h_{c}(\vartheta)$.

$\mathbf{B.5}$ Range:
$\mathcal{H}_{m}-2<\mathcal{H}\leq\mathcal{H}_{m}-1$. For the
amplitudes $0\leq h<
h_{c}(\vartheta)(2-(\mathcal{H}_{m}-\mathcal{H}))$ magnetization
is described by Eq.(\ref{Region-B-3}) and for the final interval
$h_{c}(\vartheta)(2-(\mathcal{H}_{m}-\mathcal{H}))\leq h\leq
h_{c}(\vartheta)$ we have new formula for the the magnetic moment
$\mathcal{M}(h)$
\begin{equation}\label{Region-B-5}
\left\{
\begin{array}{l} h_{c}(\vartheta)(2-(\mathcal{H}_{m}-\mathcal{H}))\leq h\leq
h_{c}(\vartheta),\\
\mathcal{M}(h)=\frac{\cos^{2}\vartheta+\delta\sin^{2}\vartheta}{8\pi}
\left(1-\frac{h}{h_{c}(\vartheta)}\right)^2.
\end{array}\right.
\end{equation}

$\mathbf{B.6}$ Range: $0<\mathcal{H}\leq\mathcal{H}_{m}-2$. For
all possible amplitudes of alternating magnetic field $0\leq h\leq
h_{c}(\vartheta)$ magnetization $\mathcal{M}(h)$ s described by
the Eq.(\ref{Region-B-5}).

Behavior of $\mathcal{M}(h)$ in the ranges $\mathbf{B.5}$ and
$\mathbf{B.6}$ is completely different form that considered above.
For these ranges there is no cusp in the distribution of magnetic
induction components in the sample and magnetization is always
positive for any interval of $h$ so that sample is totally
paramagnetic. Growth of the amplitude of alternating field
smoothly suppress magnetic moment which nevertheless remains
positive.

\section{Conclusions}

Theoretical analysis and recent
experiments\cite{Instabil3,Instabil4,Experiment} shows that the
alternating magnetic field exerts significant influence on the
static magnetic properties of anisotropic disordered high-$T_{c}$
superconductor. Switching on a sufficiently strong field
$\mathrm{\mathbf{h}}(t)$ orthogonal to the static magnetizing
field results in the complete suppression of the magnetic moment
of the sample. The reason for this suppression is that, at all
places where the alternating field penetrates, levelling of the
distribution profile of the static magnetic induction is observed.
In the other words, in the same spatial region of the sample, the
constant and alternating screening currents cannot coexist. In the
conditions when the alternating field penetrates into the entire
volume of the sample, complete suppression of the static
magnetization takes place. The nature of the magnetization
suppression, which consists in a local effect of the mutual
influence of different components of the critical current density
vector, is manifested in the anisotropic situation in a rather
peculiar way. Different components of the magnetic field penetrate
at different depths, since the are screened by critical current
densities components of quite different magnitudes. This is the
reason why the magnetization suppression is primarily caused by
the alternating filed component deeply penetrating into the
sample. As the result, the anisotropy induces a quite interesting
effect: to suppress large magnetic moment a small amplitude of the
alternating signal is sufficient. In the paper, the dynamics of
the magnetization suppression with increase in the amplitude of
the alternating filed $h$ is studied in details and results are in
agreement with that obtained in the experiments
Ref.\cite{Instabil3,Instabil4,Experiment}. It was shown that in
some cases the dependence of the moment on $h$ is nonmonotonic
and, in addition, during the suppression transition of the sample
from the paramagnetic state into diamagnetic state sometimes
occurs. All results can be interpreted within the framework of a
critical state model generalized to the anisotropic case.

\end{document}